# A new family of field-stable and highly sensitive SQUID current sensors based on sub-micrometer cross-type Josephson junctions

M Schmelz[1], V Zakosarenko[2], T Schönau[1], S Anders[1], J Kunert[1], M Meyer[2], H-G Meyer[1] and R Stolz[1]

[1] Leibniz Institute of Photonic Technology, PO Box 100239, D-07702 Jena, Germany
[2] Supracon AG, An der Lehmgrube 11, D-07751 Jena, Germany

**Abstract**
We report on the development of a new family of SQUID current sensors based on sub-micron cross-type Josephson tunnel junctions. Their low total junction capacitance permit high usable voltage swings of more than 100 µV and exceptional low noise of the SQUIDs at 4.2 K. Integrated rf-filter as well as high tolerable background fields during cool-down of up to 9.6 mT enable their highly reliable and easy use. With input coil inductances ranging from 10 nH to 2.8 µH and current sensitivities and coupled energy resolution down to 65 fA/Hz$^{1/2}$ and below 10 h, respectively, they are a versatile tool for numerous applications.

___________________________________________________________________________________

## 1. Introduction

Superconducting Quantum Interference Devices (SQUIDs) are today still the most sensitive sensors for the detection of magnetic flux. Modern thin film devices, equipped with integrated superconducting pickup coils which produce magnetic flux in the SQUID loop due to a current in this coil, are very versatile and can sense any quantity that can be converted into electrical current. Such current sensors are used i.e. for the readout of cryogenic detectors, noise thermometry, metrology or as magnetic field and gradient sensors utilizing superconducting antenna structures, i.e. in ultralow-field magnetic resonance imaging [1-4].

The most important figures of merit for SQUID current sensors are the equivalent input current noise $S_I^{1/2} = S_\Phi^{1/2}/M_{in}$ and the coupled energy resolution $\varepsilon_C = S_I L_{in}/2$. Here $S_\Phi^{1/2}$ and $M_{in}$ denote the flux noise of the SQUID and the mutual inductance of the input coil to the SQUID and $L_{in}$ the input coil inductance. Nowadays commercially available SQUID sensors with integrated input coils at 4.2 K typically offer equivalent input current noise level of about 1 pA/Hz$^{1/2}$ in the white noise region [5-7] and coupled energy resolutions below about 100 h, with h being Planck's constant [8].

By making use of additional flux transformers, increased number of turns of the input coil or by use of high inductance SQUIDs, the coupling $M_{in}$ and thus the input current noise $S_I$ can be improved. This, however, typically comes along with a degraded coupled energy resolution. It is thus difficult to design SQUIDs with excellent current noise and coupled energy resolution at the same time.

Another possibility is to decrease the total capacitance of Josephson junctions. Accordingly, the flux noise decreases, according to $S_\Phi^{1/2} = 4L_{SQ}^{3/4}C_{JJ}^{1/4}(2k_BT)^{1/2}/\beta_C^{1/4}$, where $L_{SQ}$, $C_{JJ}$ and $k_BT$ are the SQUID inductance, the total junction capacitance and the thermal energy, respectively [9]. By using sub-micrometer sized cross-type Josephson junctions [10], very low noise SQUID magnetometer or miniature SQUIDs have been developed in the past [11-14] with energy resolution approaching the quantum limit.

In this work, we report on the development of a new family of SQUID current sensors based on sub-micron cross-type Josephson tunnel junctions. Their small junction size not only results in an accordingly small capacitance, but moreover leads to a high stability against magnetic background fields during cool-down and operation [11, 15].

In Section 2 we describe the sensor design. A thorough characterization of sensor performance at 4.2 K together with theoretical estimations is given in Section 3. In Section 4 we characterize the field stability of the SQUIDs during cool-down.

## 2. Sensor Design

The presented SQUID sensors were fabricated with the IPHT cross-type Josephson junction technology [10]. With a nominal junction size of (0.8 × 0.8) µm$^2$ and a critical current density of about 1.5 kA/cm$^2$ the total junction capacitance results to about 40 fF [15].

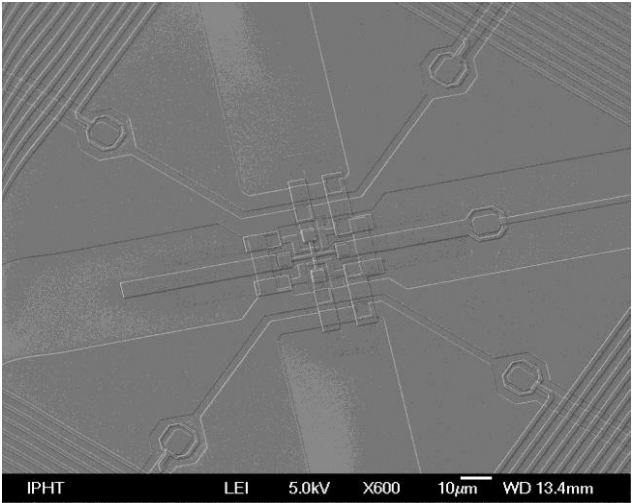

**Figure 1.** Scanning electron microscope image showing the central part of a SQUID current sensor CE1K34 with the four washers starting in the corners and parts of the input coils on top of them.

Thus, even with a moderate McCumber parameter $\beta_C$, large hysteresis-free usable voltage swings can be achieved. For the given critical current density the critical current for one Josephson junction is about $I_C \approx$ (8-10) µA at 4.2 K. For devices described within this work a shunt resistance of about 20 Ω has been selected, so the McCumber parameter $\beta_C = 2\pi I_C R^2 C_{JJ}/\Phi_0$ has been set to (0.4-0.5) to obtain smooth flux-voltage characteristics. Otherwise, especially SQUIDs with large input coil inductances tend to exhibit resonances in their flux-voltage characteristics. As we will see in Section 3 even with such moderate values of $\beta_C$ large usable voltage swings of more than 100 µV can easily be achieved.

The SQUIDs are based on a cloverleaf structure with four main washers, as proposed in [16]. With inner hole diameters between 285 and 320 µm the estimated inductances of one washer $L_W$ range from 610 to 650 pH, as listed in Table I. Inductance estimations have been carried out using Fasthenry [17]. With four washers in parallel a total SQUID inductance $L_{SQ}$ of about (160-170) pH was determined. The modulation parameter $\beta_L = 2L_{SQ}I_C/\Phi_0$ thus results to about 1.6, so washer shunts have been introduced as described in [18].

Figure 1 shows a scanning electron microscope image of the central part of the SQUID current sensor of type CE1K34 around the Josephson junctions. The four washers are starting in the corners and one can see first outer turns of the input coil on top of the SQUID washers with a linewidth and spacing of each 2 µm.

We developed a family of SQUID current sensors with integrated thin film input coils on the SQUID washer. Table I summarizes the main design parameters as well as measurement results of the SQUIDs, named CE1K$N$, with $N$ giving the number of turns in the input coil. $N$ ranges from 2 to 34 and the according input coil inductances from 10 nH to about 2.8 µH, respectively. The inverse mutual inductances $M_{in}$ of the input coils range from 1.6 µA/$\Phi_0$ to 0.1 µA/$\Phi_0$ in steps of about a factor of two. In order to enable a high overall coupling constant $k_{in}$ and a very low coupled energy resolution the input coils are directly integrated on top of the SQUID washer and a double-transformer scheme like in [19] has been omitted in our SQUIDs. For the investigated devices, the typical critical current of the thin film input coils amounts to (20-40) mA. All SQUIDs are fabricated on (2.5 × 2.5) mm$^2$ square chips, including integrated on-chip rf-filter for the input coils.

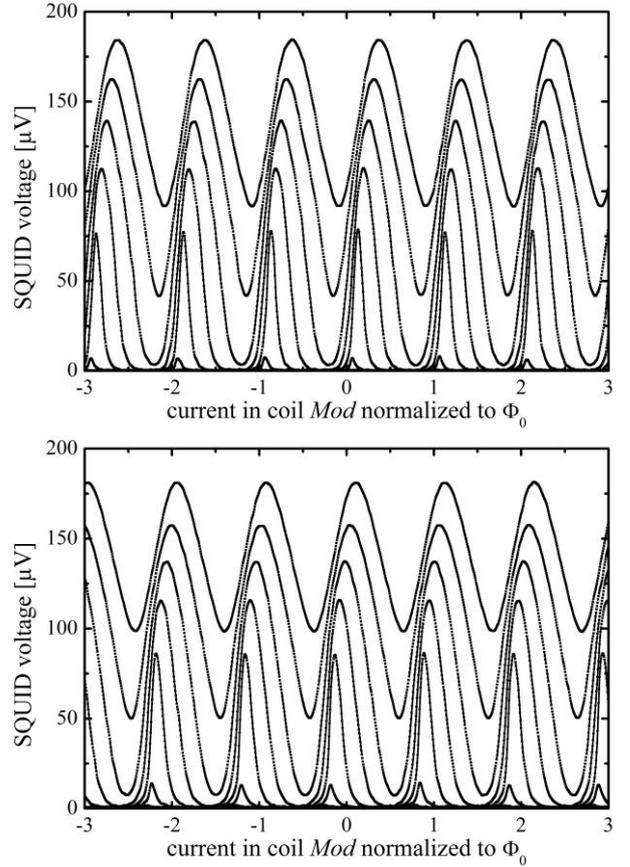

**Figure 2.** Typical set of $V$-$\Phi$ characteristics of current sensor SQUID CE1K2 (top) and CE1K34 (bottom). Bias current increases in steps of 2 µA between individual characteristics.

The SQUIDs comprise two feedback options: Coil *Mod* produces a flux directly in the SQUID via inductive coupling and is used for flux locked loop operation. The *Fb* coil is a part of a flux transformer connected in series with the input coil with possibly smallest coupling directly to the SQUID inductance. It produces a flux in the superconducting input circuit and can be used as a current feedback for current locked loop. While coil *Mod* has an inverse mutual inductance of about 19 µA/$\Phi_0$ for *Fb* coil this accounts to about 15 µA/$\Phi_0$ assuming a superconducting short on the input coil.

**Table 1.** Characteristic parameters of SQUID current sensors CE1K*N* measured at 4.2 K.

| Device name | CE1K2 | CE1K4 | CE1K8 | CE1K17 | CE1K34 |
|---|---|---|---|---|---|
| Winding number of input coil $N$ | 2 | 4 | 8 | 17 | 34 |
| $L_W$ [pH] | 650 | 650 | 650 | 610 | 610 |
| $L_{SQ}$ [pH] | 170 | 170 | 170 | 160 | 160 |
| $L_{in}$ [nH] | 10.7 | 44 | 174 | 723 | 2860 |
| $1/M_{in}$ [µA/$\Phi_0$] | | | | | |
|    design | 1.60 | 0.80 | 0.40 | 0.20 | 0.10 |
|    measured | 1.57 | 0.79 | 0.40 | 0.20 | 0.10 |
| $M_{fb}$ [nH] | 1.5 | 2.2 | 4.5 | 9.8 | 20.3 |
| $k_{in}$ | 0.98 | 0.96 | 0.95 | 0.96 | 0.97 |
| Intrinsic flux noise $S_\Phi^{1/2}$ [µ$\Phi_0$/Hz$^{1/2}$] | 0.55 | 0.55 | 0.55 | 0.58 | 0.65 |
| Input current noise $S_I^{1/2}$ [pA/Hz$^{1/2}$] | 0.86 | 0.43 | 0.21 | 0.12 | 0.065 |
| Energy resolution: | | | | | |
|    $\varepsilon$, uncoupled [h] | 5.8 | 5.8 | 5.8 | 6.8 | 8.5 |
|    $\varepsilon_C$, coupled [h] | 6.0 | 6.3 | 6.4 | 7.3 | 9.1 |

## 3. Sensor performance

As the Josephson junctions in the IPHT cross-type technology are defined by the overlap of two narrow strips of different superconducting layers, any parasitic capacitance in parallel to the junction is avoided. Together with submicron junction dimensions, the resulting small total junction capacitance allows large shunt resistor values while preserving non-hysteretic SQUID characteristics. The transfer function $V_\Phi$ and usable voltage swing strongly increase as they scale with $1/\sqrt{C_{JJ}}$ [15]. For devices described within this work we typically achieve usable voltage swings of (100-150) µV at 4.2 K. Figure 2 shows typical sets of flux-voltage characteristics of devices CE1K2 and CE1K34 where bias currents have been increased in steps of 2 µA. The visible flux-shift for increasing bias-currents is due to an asymmetric bias-injection in the SQUID inductance, leading to a reduction of input current noise contribution of the SQUID electronics [20]. One should point out that the measured flux-voltage characteristics did not show significant differences between SQUIDs with differing number of input coil turns. Device characterizations as well as all subsequent noise measurements were carried out at 4.2 K with the SQUIDs immersed in liquid helium inside a lead and µ-metal shield.

For devices with further increased usable voltage swings due to increased critical current density of the junctions we observe a usable voltage swing of more than 200 µV. Although these devices show somewhat larger McCumber parameter of $\beta_C \approx 0.65$, we measured slightly higher equivalent white flux noise levels of i.e. 0.6 µ$\Phi_0$/Hz$^{1/2}$ for type CE1K2. They moreover tend to exhibit resonances in their characteristics and we thus limit the McCumber parameter to about (0.4-0.5) for devices described in this work to obtain smooth flux voltage characteristics and which offer a little lower flux noise.

Noise measurements were carried out using a directly-coupled low-noise SQUID electronics from Supracon AG with an input voltage and input current noise of $S_V^{1/2} = 0.35$ nV/Hz$^{1/2}$ and $S_I^{1/2} = 6.5$ pA/Hz$^{1/2}$, respectively [5]. Even so, in a single stage configuration the typically measured white flux noise of about 1 µ$\Phi_0$/Hz$^{1/2}$ is dominated by contributions from the input voltage noise of the room-temperature electronics $S_V^{1/2}/V_\Phi$.

To exploit the intrinsic noise of the SQUIDs, we used a two-stage readout configuration with a second SQUID as a low noise preamplifier [21]. Figure 3 shows the simplified circuit diagram of the two stage measurement setup. The SQUID SQ$_1$ under investigation has been voltage biased with $R = 1\,\Omega$ and the amplifier SQUID was operated as an ammeter. It thus senses the current change due to an external signal to SQ$_1$. Feedback from a commercial low-noise directly-coupled flux locked loop electronics [5] was applied to SQ$_2$. Moreover, we introduced a choke inductor preventing the coupling of high-frequency noise between the two SQUIDs [22].

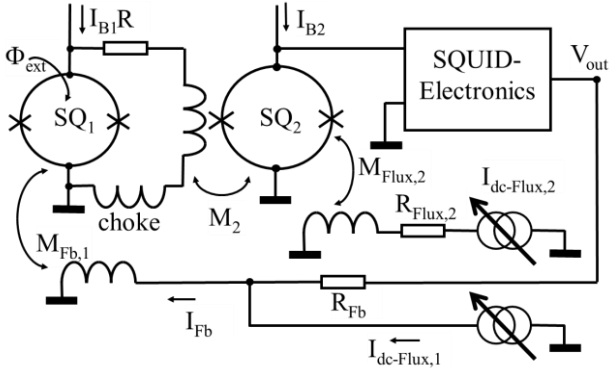

**Figure 3.** Simplified circuit schematic for the two-stage noise measurements: $SQ_2$ acts as an ammeter and senses the current change due to an external signal to $SQ_1$. The equivalent flux noise, as shown in Figure 4, was calculated from the measured voltage noise at the output of the SQUID electronics using the measured overall transfer function.

The electronics output voltage has been recorded with an HP 3565 spectrum analyzer with a maximum bandwidth of 100 kHz. Representative flux noise spectra of SQUIDs of type CE1K8 and CE1K34 are shown in Figure 4.

For SQUIDs CE1K$N$ with N ranging from $N = 2$ to 8 white flux noise levels of 0.55 $\mu\Phi_0/Hz^{1/2}$ has been measured, corresponding to an energy resolution of about 6 h, as listed in Table 1. For SQUID CE1K8 with an input inductance of about $L_{in} = 174$ nH this corresponds to an input referred current noise $S_I^{1/2} = 210$ fA/Hz$^{1/2}$.

For SQUIDs with increased input coil inductance, due to an increased number of turns of the input coil, the measured white flux noise slightly increases. For SQUIDs with $N = 34$ white flux noise levels of about 0.65 $\mu\Phi_0/Hz^{1/2}$ have been measured, corresponding to an energy resolution and input referred current noise of about 8.5 h and 65 fA/Hz$^{1/2}$, respectively.

The measured white flux noise agrees very well with theoretical predictions resulting from $S_\Phi^{1/2} = 4L_{SQ}^{3/4}C_{JJ}^{1/4}(2k_BT)^{1/2}/\beta_C^{1/4}$, with $C_{JJ} = 40$ fF, $T = 4.2$ K and $\beta_C \approx 0.5$. Moreover, the observed increase in the white flux noise level for large input coil inductances is qualitatively consistent with estimations from [23]. Here, the input coil introduces a parasitic capacitance $L_p$ across the SQUID inductance thereby degrading the energy resolution. For devices described in this work this increase, however, shows a much weaker dependence than those reported above. Accordingly, even devices with $L_{in} \approx 2.8$ $\mu$H and thus large $L_p$ show less than twice the energy resolution of low input coil inductance devices.

The reported SQUIDs exhibit very low white input current noise levels down to 65 fA/Hz$^{1/2}$, which is lower than previously reported comparable integrated thin-film current sensor SQUIDs [8, 24, 25]. A further increase of $N$ by i.e. the use of smaller linewidth input coils [16] or combining such a device with an additional superconducting thin-film transformer as reported in [26] may even further improve the input current noise and may be integrated onto the same chip.

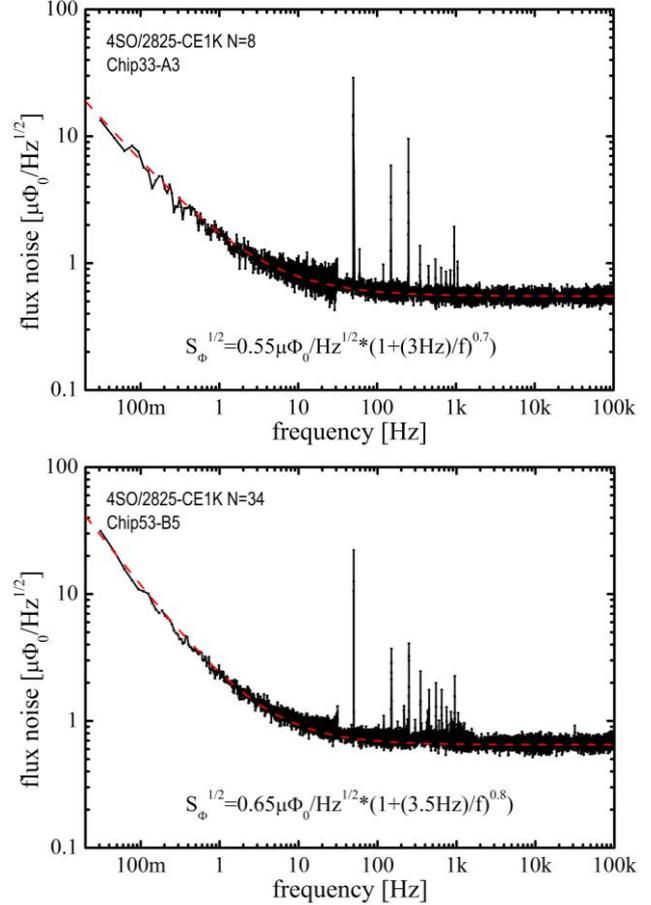

**Figure 4.** Equivalent flux noise spectra of current sensor SQUID CE1K8 (top) and CE1K34 (bottom) as measured in a two-stage configuration and as explained in the text. The white flux noise amounts to 0.55 $\mu\Phi_0/Hz^{1/2}$ and 0.65 $\mu\Phi_0/Hz^{1/2}$ for CE1K8 and CE1K34, respectively. The red line shows the fit according to the expression given in the figures.

As we achieved a tight coupling with $k_{in} \approx 0.95$-0.98 between the input coil and SQUID at the same time the presented SQUIDs exhibit a very low coupled energy resolution of i.e. 6.0 h for SQUIDs of type CE1K2. As given in Table 1 all presented devices show a coupled energy resolution below 10 h in the white noise region! This represents roughly a factor of four improvement compared to e.g. [8].

Coupling constants, as given in Table 1, have been calculated as $k_{in} = M_{in}/(L_{in}L_{SQ})^{-1/2}$, where $M_{in}$ is the measured mutual inductance between the input coil and the SQUID and $L_{in}$ estimated as $L_{in} = L_{strip} + 4N^2 L_W$, with $L_{Strip}$ being the stripline inductance [27].

For frequencies below about $f = 1$ Hz the noise is dominated by critical current fluctuations in the Josephson junctions, with a $1/f^\alpha$ dependence with $\alpha \approx (0.7\text{-}0.8)$ at 4.2 K. Assuming a typical current sensitivity of LTS SQUIDs of $M_{dyn} = (\partial\Phi/\partial I_B) \approx (1\ldots2)\cdot L_{SQ}$ [20] the empirical based formula for critical current fluctuations in $AlO_x$ based Josephson tunnel junctions [28] shows very good agreement for the measured magnitude of flux noise at 1 Hz and 4.2 K.

One should moreover point out, that, as shown in Figure 4, SQUID CE1K8 exhibits a magnitude of flux noise at 1 and 10 Hz of 1.7 $\mu\Phi_0/Hz^{1/2}$ and 0.8 $\mu\Phi_0/Hz^{1/2}$, respectively. With a SQUID inductance $L_{SQ} = 170$ pH this corresponds to an energy resolution of 55 h and 12 h, respectively.

**4. Magnetic field stability**

Besides the significant improvements in input current noise and coupled energy resolution of the SQUIDs, a high stability against dc magnetic background fields is expected due to the small linewidth of superconducting structures at and close to the Josephson junctions. In this regard, field stability denotes the magnetic field for vortex trapping in or close to the Josephson junctions that would affect their critical currents.

For the measurement we set up the SQUID chip in the center of a solenoid magnet assembled to our dipstick with the applied magnetic field perpendicular to the chip surface. The SQUID was biased with constant current so that the measured voltage swing was maximal. We swept the current through coil *Mod* and measured the flux-voltage characteristics. The chip has been heated up by an on-chip resistor while a constant magnetic field was applied. After cooling down the flux-voltage characteristics has been recorded while the magnetic background field was kept constant. The trapped flux in the junctions was recognized from a shift of flux-voltage characteristics along the voltage axis resulting from a change of the junction critical current. Repeating this procedure for each value of the magnetic background field and stepwise increasing the magnitude of the background field, we obtained a histogram-like distribution of flux trapping probability in the junctions.

The presented SQUIDs have proven to withstand magnetic background fields with an average of 9.6 mT during cool-down with a typical crossover width from zero to 100% flux trapping probability of about 0.1 mT. While the minimum measured critical field for flux trapping was as high as 8.4 mT for all our tested devices there have been some outliers with a critical field of more than 12.3 mT, which was the maximum field amplitude our setup allows for. Besides this, we do not observe an influence of the number of turns in the input coil on the measured critical field for vortex trapping in the junctions.

Our measurement results roughly agree with theoretical estimations given in i.e. [29, 30]. Therein, the field for vortex trapping in type-II superconducting strips is given as $B_0 \propto \Phi_0/w^2$, with $w$ being the strip width. For the used junction dimension of $(0.8 \times 0.8)$ $\mu m^2$ $B_0$ results to about 5.1 mT [29].

We currently attribute the increased critical fields for vortex trapping compared to previous investigations on multiloop SQUID magnetometer [15] to the fact that the actual junction dimension may be slightly reduced during fabrication and to the particular design of the SQUID. Further studies on different SQUID designs may help to understand this phenomenon and probably to further increase this value.

The presented devices outperform their counterparts based on conventional window-type junctions [8]. Their improved reliability is expected to expand their application range. A further reduction in junction dimension should further enhance the magnetic field stability.

**5. Conclusions**

We developed a new family of field-stable and highly sensitive SQUID current sensors based on submicron cross-type Josephson tunnel junctions. Their input coil inductances vary between 10 nH and 2.8 $\mu$H. Due to the small total junction capacitance the SQUIDs feature a large usable voltage swing of typically (100-150) $\mu$V and very low noise. Due to the tight coupling of the input coil they exhibit white input current noise level down to $S_I^{1/2} = 65$ fA/Hz$^{1/2}$ and at the same time coupled energy resolutions below 10 h. Even at low frequencies they offer very low energy resolutions of 55 h and 12 h at 1 and 10 Hz, respectively. Integrated rf-filter as well as their ability to cool down without flux trapping in magnetic background fields of up to 9.6 mT ensures their reliability and easy use as well as expanding their possible application range. They are thus a versatile tool for numerous applications, like i.e. the readout of cryogenic detectors or as magnetic field and gradient sensors.

**References**

1. Kraus Jr R, Espy M, Magnelind P and Volegov P, *Ultra-Low Field Nuclear*